**Effects of Characteristic Length Scales on the Exciton Dynamics in Rubrene Single Crystals**

*Björn Gieseking\*, Teresa Schmeiler, Benjamin Müller, Carsten Deibel, Bernd Engels, Vladimir Dyakonov, and Jens Pflaum*

[\*]     Prof. J. Pflaum, B. Gieseking, T. Schmeiler, B. Müller, Dr. C. Deibel, Prof. V. Dyakonov
Experimental Physics VI, Julius-Maximillian University,
Würzburg, D-97074 (Germany)
E-mail: Bjoern.gieseking@physik.uni-wuerzburg.de

        Prof. J. Pflaum, Prof. V. Dyakonov
Bavarian Center for Applied Energy Research e.V. (ZAE Bayern),
Würzburg, D-97074 (Germany)

        Prof. B. Engels
Institute for Physical and Theoretical Chemistry, Julius-Maximillian University,
Würzburg, D-97074 (Germany)

The transport of optically excited states, also referred to as excitons, defines a critical step in the light-to-charge carrier conversion in organic materials.[1] To overcome limitations by the short diffusion lengths of singlet excitons, the harvesting of triplet excitons generated via singlet fission has been suggested as a possible approach to increase the conversion efficiency in organic photovoltaic cells.[2,3] Furthermore, upon further miniaturization of up-to-date organic electronic devices the correlation between exciton transport and morphological length scales becomes increasingly important as the structural coherence significantly affects the photo-physical processes and their time-evolution.[1, 4, 5] Here, we demonstrate the influence of spatial confinement on the excitonic states and their dynamics in the prototypical organic semiconductor rubrene. By its unique property of providing morphologies with different characteristic length scales we observe direct structural influence on the temperature dependent Photoluminescence (PL) spectra and on the relaxation processes on $10^{-12}$ s time scales already for excitation volumes with µm lateral extensions. Our results highlight the role of the local environment which affects the exciton dynamics in organic semiconductors and which has to be considered upon tailoring the microscopic morphology of organic opto-electronic devices.

Besides its remarkable photophysical behavior, characterized e.g. by singlet fission in combination with extended triplet lifetimes (µs) and extended diffusion lengths (µm),[6-8] rubrene offers the possibility of different, well-controllable structural morphologies in combination with sufficiently high PL quantum yields. In thin films without further surfactants rubrene grows x-ray amorphous due to an activation barrier of 210 meV[9] required for planarization of its conjugated backbone. In contrast, single crystals grown by vapor sublimation under streaming carrier gas show an exceptional structural quality characterized by µm-sized crystalline facets as well as lateral extensions of several millimeters.[10] Adjusting the conditions during growth allows for a broad morphological variety of microstructures in-between, ranging from spatially defined microcrystals to self-organized pyramidal surface structures,[11] all of which imposing different boundary conditions for photogenerated excitons.

A recent study on the directionality of the steady state absorption and emission of rubrene suggests that variations of the PL spectra detected on microstructures originate from the anisotropic nature of emission.[12] Though the relative intensities can be explained by this approach it is insufficient to account for the temperature dependent optical phenomena and their temporal evolution on short time scales reported in this work.

Following pulsed laser excitation we probed the temperature dependence of the photoluminescence decay dynamics of three types of rubrene morphologies with different structural hierarchies - bulk crystals, microcrystals and amorphous thin films (**Fig. 1a** and **Fig. 4** and **5** in supplementary information). By variation of their inherent length scales these samples exhibit an increasing degree of spatial confinement for photogenerated excitons, which directly affects the time-integrated PL spectra (**Fig. 1b** and **Fig. 6** in supplementary information). Compared to the bulk spectrum, the high-energy peak (2.16 eV) is significantly enhanced in the microcrystal PL spectrum. This peak has been associated either with a charge transfer state[6] or a coherent excitonic state[11, 13, 14] whereas the lower energetic peaks (2.02, 1.88 and 1.74 eV) are attributed to emission from a localized molecular excitonic species with a vibronic progression of 140 meV.[15,16] However, in our data we find no indication for the coexistence of two independent emitting species in rubrene and therefore consider the radiative decay originating from just one singlet excitonic state with different polarizations and being the only contribution to the detected PL signal. The enhancement of the 2.16 eV peak in the microcrystals can therefore be related to enhanced emission polarized parallel to c direction (M-polarization) from the side facets of the confined crystalline volume and to the reduced self-absorption within the microstructures.[12]

The shape of the microcrystal PL is very similar to that of the amorphous film, which can be explained by the fact that the directionality of emission and the enhanced out-coupling at the crystal edges is compensated by the random orientation of molecules in the amorphous film.[12] However, the peak width of the latter is broadened due to the enhanced energetic disorder. As a result of this disorder excitations are preferentially localized on molecular length scales and, in analogy to free charge carriers, expected to exhibit a reduced mobility compared to crystalline samples.[17] Furthermore, the triplet quantum yield by inter-system crossing is assumed to be low in these samples in accordance to observations on rubrene in solution. [18, 19]

Optical excitation of the two crystalline structures (bulk and microcrystal), in contrast, leads to creation of more mobile excitons. At 4 K a lower limit for the exciton diffusion constant of 0.2 cm$^2$/s has been deduced which is in line with values reported for other polyaromatic single-crystals of up to 10 cm$^2$/s at cryogenic temperatures.[11,20] Providing long-range molecular ordering, excitons are therefore able to travel large distances before relaxing via radiative or non-radiative processes such as singlet fission[7] or quenching at defects. As a consequence, in spatially confined crystalline microstructures low-energy trapping sites induced by local disorder at the boundaries impose a significant influence on the decay dynamics, if reached by excitons.[21] We point out that we measured an ensemble of crystals with micrometer sizes and that for large crystals also boundaries between crystalline subdomains might have to be taken into account.

Appraising this microscopic model by temperature dependent PL studies on the various samples, the amorphous rubrene film initially shows a PL increase upon cooling from room temperature followed by a continuous drop of intensity below 150 K (see **Fig. 2a** and **b**). The primary rise of the PL signal with decreasing temperature can be ascribed to a suppression of thermally activated, non-radiative decay processes. The deduced barrier of $\Delta E_3 = 134$ meV is of the order of molecular vibrational energies and thus points at the local character of this decay channel. Below 150 K the lack of thermal activation energy promotes additional non-radiative decay channels, e.g. by static impurity quenching.

In contrast, cooling down the two crystalline morphologies to 8 K leads to a pronounced monotonous increase of the detected PL intensity (see **Fig. 2a** and **b**). In accordance to the thermal PL behavior of the amorphous film, the absolute change can be explained by the existence of a thermally activated non-radiative decay mechanism. For bulk and microcrystals, however, we deduce different barrier heights of $\Delta E_1 = 44$ and $\Delta E_2 = 25$ meV, respectively (see also **Table 1**). These values are in agreement with activation energies observed for other long-range ordered molecular systems such as α-PTCDA single crystals[22].

By complementary analysis of the time-resolved PL spectra we are able to identify the fundamental decay channels and their dynamics dominating the transient spectroscopic behaviour of excitons and, as demonstrated, being in tight correlation with the respective degree of spatial confinement. For the bulk crystal three thermally activated non-radiative decay channels exist (**Fig. 3a**) and have been assigned by their respective decay times $\tau = \gamma^{-1}$ [23]:

$$\gamma = \gamma_r + \gamma_{nr}^{(1)} e^{-E_1/kT} + \gamma_{nr}^{(2)} e^{-E_2/kT} + \gamma_{nr}^{(3)} e^{-E_3/kT}. \qquad (1)$$

Here $\gamma_r$ describes the decay rate of the radiative decay channel, $\gamma_{nr}^{(i)}$ of the non-radiative decay channels and $E_i$ the energies of the corresponding activation barriers. At this point, it is important to note that because of the linear relation between PL intensity and laser excitation power we can exclude bimolecular processes as an alternative decay channel as well as delayed fluorescence by triplet fusion.[8]

At room temperature the PL decay of the bulk crystal is dominated by a non-radiative decay channel with a time constant of 20 ps ($1/\gamma_{nr}^{(1)}$), which is consistent with the picosecond component found by Stöhr et al. (**Fig. 3b**).[11] An additional exponential decay with a time constant of 100 ps is present ($1/\gamma_{nr}^{(2)}$), however yielding a much smaller contribution to the PL decay. The 20 ps relaxation channel is in excellent agreement with the reported rise time of the triplet absorption band due to for singlet fission in rubrene single crystals and the estimated energetic barrier of 45 meV matches the offset between the energy of the excited singlet state $S_1$ and the summed-up triplet energy of $T_1 + T_1$.[18] An additional ~2 ps component reported by Ma et al. could not be resolved in our measurements, presumably due to limitations by our setup. In microcrystals this fission process cannot be identified. There, the PL decay is rather governed by the 100 ps decay component with a corresponding activation energy of 25 meV deduced by the temperature dependence of the integrated PL spectrum. This observation can be directly correlated with the spatial confinement of optical excitations in the microcrystal volume and points towards a surface related quenching mechanism originating from the local disorder at the boundaries and leading to an effective time constant $\gamma_{nr}^{(2)}$. A time constant of 100 ps at room temperature has also been observed by Tao et al.[24] and explained by formation of a polaronic absorption band upon exciton-polaron conversion. While the elementary dissociation mechanism could not be identified the results point towards a material-inherent channel which is in agreement with the assignment of a surface related process. Another possible loss scenario originates by the conversion of singlet excitons into excitonic dark states at the interface. Theoretically, such a process related to disorder on molecular length scales has been predicted for various polyaromatic hydrocarbons.[21] The local character of this decay channel is supported by the absence of lattice vibrational modes in this energy range.[25] As relaxation via triplet formation requires at least twice the activation energy of the 100 ps decay channel and as this energy is expected to be further enhanced by the reduced polarizability at microcrystal boundaries, singlet fission is effectively suppressed in confined µm-sized microcrystals. This microscopic picture is supported by the elongated exciton lifetime by a factor of five.

The third decay channel with a time constant of around 0.5 ns observed at room temperature, though much less pronounced, is in range of the effective radiative lifetime of the singlet excitons (intrinsic radiative lifetime: 16.5 ns[26]) and results from inhomogeneities within the organic crystal acting as quenching sites during exciton migration ($1/\gamma_{nr}^{(3)}$). The influence of this third channel is negligible for the photophysics of bulk single crystals but becomes relevant for exciton motion in microcrystal geometries of restricted dimensions.

Due to the localized character of the excitations, the decay of the PL signal for the amorphous film can be described by assuming solely the aforementioned reduced radiative lifetime, i.e. the non-radiative decay channel $\gamma_{nr}^{(3)}$. Interestingly, as soon as the accessibility of interfaces is provided to an amorphous film of comparable thickness and surface flatness, e.g. by grain boundaries or dislocations, the decay dynamics change and become dominated by an additional rate coinciding with $\gamma_{nr}^{(2)}$ (**Fig. 7** in supplementary information). This observation again approves this channel to be associated with interface quenching. An additional decay time of about 2.2 ns has been reported by Piland et al.[27] for the spectrally integrated PL transients in amorphous rubrene films. However, this component which was assigned to singlet fission cannot be detected in our measurements presumably due to the 2 ns time window of the utilized experimental setup. The fact that this time constant for singlet fission is larger by two orders of magnitude compared to studies by Ma et al.[12] as well as by us highlights the influence of structural disorder on the decay dynamics in amorphous rubrene. Reducing the temperature of bulk and microcrystal leads to an increase of the average radiative lifetimes of the excitonic states as the decay via thermally activated non-radiative channels is continuously hampered. Between 170 and 120 K, the picosecond decay channels completely freeze out for bulk and microcrystal samples as the population density of phonons is strongly diminished. Therefore, despite its higher activation energy the $\gamma_{nr}^{(3)}$ channel resembles the only non-radiative relaxation pathway for excitons at lower temperatures which cannot be explained by the mere anisotropy of the optical behaviour.

The decay dynamics in this temperature regime are comparable to those of the amorphous sample: With decreasing temperature, i.e. decreasing exciton diffusion, the probability of localization at quenching sites is reduced resulting in the observed increase of the average excitonic lifetimes. These lifetimes are in the order of a few ns in agreement with literature,[8, 11, 13] but due to our experimental resolution can only be determined within a time-frame of 2 ns. Nevertheless, it can be concluded that due to the enhanced influence of interfaces the observed exciton lifetime for the microcrystals become smaller.

In summary, we have demonstrated that the inherent length scales affect the excitation dynamics in rubrene samples with different morphologies and we can ascribe their temperature- and time-dependent PL behavior to the presence of three fundamental decay mechanisms and their respective predominance in the various morphologies. For single crystals we conclude that already the spatial confinement on micrometer dimensions, which is typical for organic thin film devices, has a massive influence on the temporal evolution of the PL. While the non-radiative decay in the rubrene single crystals is dominated by thermally activated singlet fission at a time constant of 20 ps and an activation barrier of 44 meV, this channel is effectively suppressed in microcrystal structures due to the presence of an additional decay mechanism related to the boundaries and with only 25 meV activation energy. As a result for microcrystals, excitons reaching the grain boundaries are trapped by surfaces states and transferred to dark states or dissociated into polarons within 100 ps. A third relaxation channel within the nanosecond time regime becomes relevant for long range ordered rubrene stacks at very low temperatures. The local character of this decay path is indicated by its coincidence in time and energy with the only relaxation process observed for amorphous rubrene thin films. Our findings clearly elucidate the effects by the local environment on the dynamic photophysical behavior of organic semiconductors. With respect to spatial dimensions of organic thin film devices, this correlation and the effects emerging by confinement have to be considered in the ongoing miniaturization and optimization of organic opto-electronic devices, such as photovoltaic cells.

Experimental Details:

All experimental procedures are described in the Supplementary Information.

*Acknowledgements*
The authors thank Prof. Martin Kamp (Würzburg University) for his assistance on TEM measurements and Stephan Hirschmann (Stuttgart University) for material purification. Financial support by DFG within Research Unit (FOR 1809) "Light induced dynamics in molecular aggregates" and within the contract INST 93/623-1 FUGG is acknowledged.

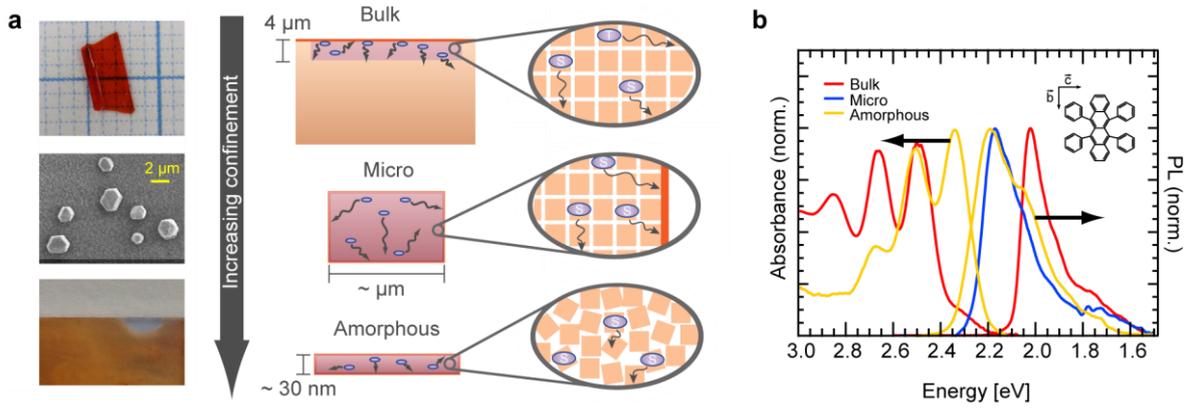

**Figure 1. a** Pictures of different samples and related morphologies of the excitation volumes schematically illustrating the effect of spatial confinement on the created excitons. **b** Time-integrated PL spectra of rubrene bulk and microcrystals as well as of a thin amorphous film together with bulk crystal and amorphous film absorption. The inset shows the molecular structure of rubrene together with the crystallographic directions.

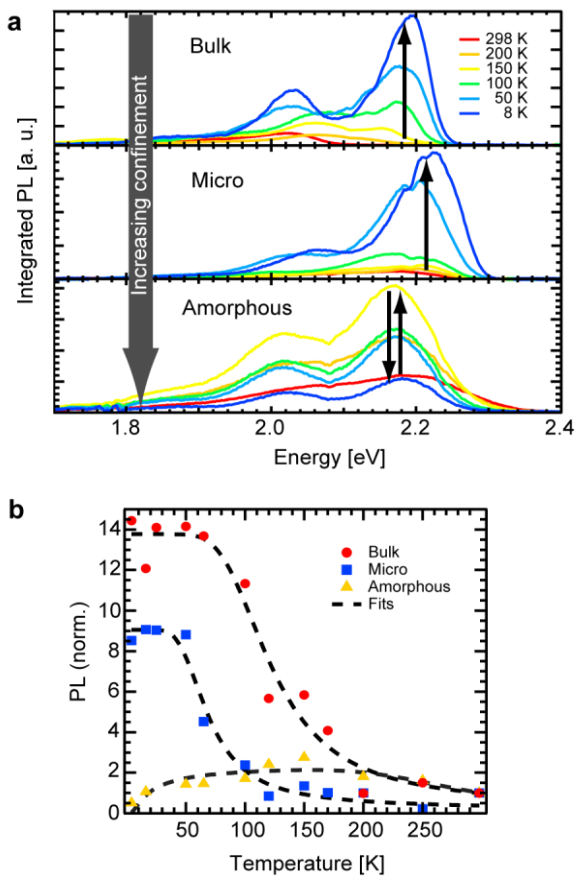

**Figure 2. a** Integrated PL intensity spectra exhibiting a dramatic increase of PL signal with decreasing temperature. For the two crystalline morphologies, the first spectra (298 and 250 K for bulk; 298-150K for micro) were multiplied by a factor of six and two, respectively. **b** Changes of the PL spectra normalized to the values at 298 K as a function of temperature. Both crystalline samples exhibit a PL increase upon reducing the temperature and saturation at low temperatures. Contrary, the PL signal of the amorphous film, after initial increase, is reduced for temperatures below 150 K.

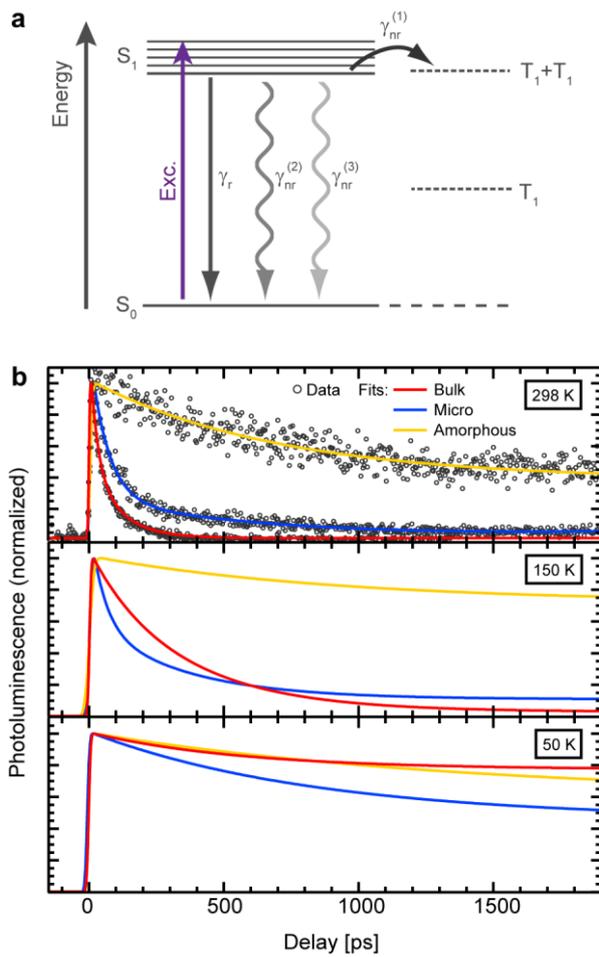

**Figure 3. a** Energy level scheme showing the possible relaxation pathways for singlet excitons in rubrene bulk crystals. **b** Fits of the transient PL data taken at 2.16 eV for all three samples at room temperature, 150 and 50 K. The temperature dependent decay indicates the presence of thermally activated non-radiative decay channels. For the sake of clarity for low temperatures only the fit transients are shown.

| Sample type | $1/\gamma_r$ | $1/\gamma_{nr}^{(1)}$ | $\Delta E_1$ [meV] | $1/\gamma_{nr}^{(2)}$ | $\Delta E_2$ [meV] | $1/\gamma_{nr}^{(3)}$ | $\Delta E_3$ [meV] |
|---|---|---|---|---|---|---|---|
| Bulk | Few ns | **20 ps** | 44 | 100 ps | 25 | ~ns | 134 |
| Micro | Few ns | --- | | **100 ps** | 25 | ~ns | 134 |
| Amorphous | Few ns | --- | | --- | | **~ns** | 134 |

**Table 1.** Parameters determined from the analysis of the temperature dependence of the steady-state and transient PL spectra. The values refer to the energetic barriers and the lifetimes (inverse rates) as well as the associated decay mechanisms for the different rubrene samples studied in this work. Lifetimes in bold letters correspond to the dominant decay channel at room temperature.

# SUPPLEMENTARY INFORMATION

**Sample Preparation**

Rubrene powder was purchased from Sigma Aldrich and further purified by gradient sublimation. Single crystals were then grown by horizontal physical vapor transport (PVT) [28] in a constant stream of high-purity 6N nitrogen gas. The source temperature during growth was set to 280 °C and nucleation took place in a distance of approximately 10 cm from the source. Needle-and platelet-like crystals were obtained with dimensions of several millimeters and thicknesses in the range of 100 - 300 µm. The achieved surfaces of the platelet-like crystals correspond to the (001)-rubrene crystal plane. Amorphous rubrene films were grown by a fast evaporation process under high vacuum conditions on a sapphire substrate. At an evaporation rate of 0.5 nm/s films of 30 nm thickness were grown. During evaporation the layer thickness was measured with an oscillating quartz crystal and afterwards confirmed by X-ray reflectivity measurements. For the microcrystals, rubrene powder was dissolved in tetrahydrofuran (THF) at a concentration of 1 mg/ml [29]. The solution was stirred for 10 minutes and then rapidly mixed with distilled water. Upon this process micro-particles were formed instantaneously. After stirring and ultrasonicating the solution for 10 minutes, amounts of several microliters were deposited on a glass substrate and kept drying at RT. Finally heating the substrate for several hours at a constant temperature of 180 °C the microcrystals were generated.

**Sample Characterization**

X-ray diffraction measurements of bulk crystal and amorphous film were carried out in Bragg-Brentano geometry with a XRD diffractometer model XRD 3003 T/T (GE Inspection Technologies) using Cu-Kα radiation at a wavelength of 1.54 Å. Due to the low particle density, characterization of the precipitated microcrystals was performed by a scanning electron microscope REM Ultra-Plus (Zeiss) using beam energies between 2 and 20 keV. To analyze the absorption properties of the single crystals spectra were detected by means of a Lambda 950 UV/VIS spectrometer (Perkin Elmer). The amorphous film was measured with a V630 UV-VIS spectrophotometer (Jasco). As the samples absorb in the visible spectral range, investigations were restricted to a wavelength range of 300 nm -800 nm (3.0 -1.6 eV). Transmission electron microscopy (TEM) studies on the micro-crystals were carried out with a FEI-Titan$^{TM}$ TEM with beam energies of 300 keV. For this, the microcrystals were deposited and then annealed on a small copper grid. For the time-resolved PL measurements, the output of a Ti:Sa oscillator (Spectra Physics, 100 fs, 800 nm) was frequency doubled and focused onto the sample, which was mounted inside a liquid helium cryostat, using a fluence of 2.6 nJ/cm$^2$. The PL was spectrally dispersed by a spectrograph and detected with a C 5680 streak camera (Hamamatsu). The temporal resolution of the setup described was 8 ps and the detection window limited to 2ns.

**Analysis of Time-integrated Spectra**

The values for the three different samples at a given temperature correspond to the area under the corresponding PL spectra. The analysis of the temperature dependence of the time-integrated spectra was carried out assuming one dominant thermally activated non-radiative decay channel: $\gamma = 1/\tau = \gamma_r + \gamma_{nr}\, e^{-\Delta E/kT}$, while the photoluminescence efficiency can be expressed as $\eta = \gamma_r/\gamma$ [23].

**Structural Characterization of Rubrene Samples**

In order to prove the amorphous character of the rubrene films we performed x-ray diffraction measurements and compared them to the results for the rubrene single crystal (**Fig. 4**). No

Bragg peaks could be found for the film confirming its amorphous character while the measurement of the single crystal clearly indicates the good crystalline quality of the analyzed sample. The spectrum of the smooth amorphous film exhibits Kiessing oscillations, which were used to determine the thickness of the layer. The size and shape of the microcrystals were deduced from the analysis of SEM pictures and the crystalline character was confirmed by carrying out transmission electron microscopy measurements (**Fig. 5**).

**Integrated PL of Crystalline Samples**

**Fig. 6** shows the comparison of the time-integrated PL spectrum of the microstructures with the ones of two bulk single crystals - one with a very smooth (bulk 1) and one with a rougher surface (bulk 2). The observed effective shift of the PL spectrum of the microcrystals is due to an enhanced relative contribution of the high-energy PL peak to the spectrum and not to the blueshift of the whole spectrum.

**Amorphous Films With Different Amount of Interfaces**

Preparing amorphous rubrene films of similar thickness and surface flatness but with different amounts of interfaces results in completely different decay kinetics at room temperature (see **Fig. 7a**). While the dynamics of the film on sapphire (squares) can be described by a monoexponential decay at a time constant of about 0.5 ns, the decay of the sample on glass (circles) is biexponetial with an additional constant of 100 ps. The time-integrated spectra of both samples are however almost identical (inset in **Fig. 7a**). This different decay behaviour is assumed to be a consequence of the different sample topographies as observed in AFM measurements (**Fig. 7b**). One can clearly see an increase of surface area introduced by dislocation boundaries for the amorphous film on glass while the film on sapphire only shows smooth plateaus.

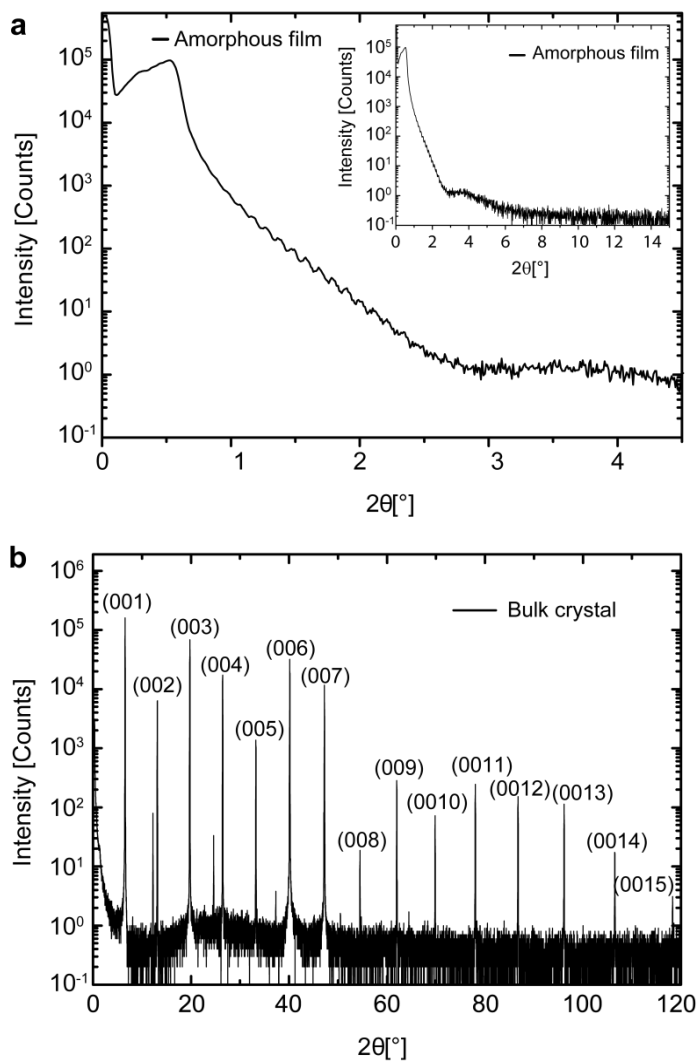

**Figure 4. a** X-ray diffraction measurements of a thin amorphous rubrene film (above) exhibiting Kiessig oscillations. The inset shows the whole Bragg spectrum where no peaks are detectable. These results confirm the x-ray amorphous character of the film. **b** The spectrum of a rubrene single crystal is displaying the high crystalline quality of the sample.

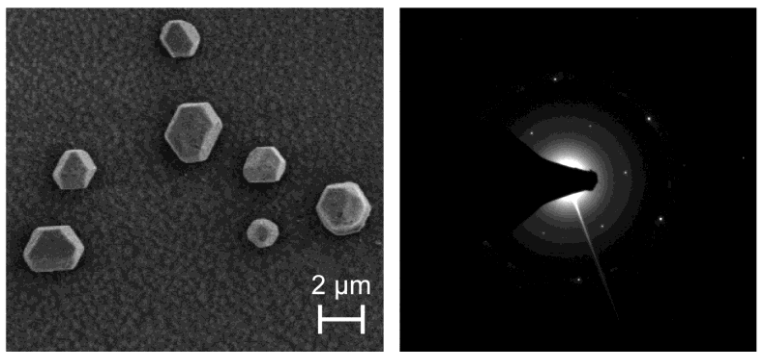

**Figure 5.** The left picture shows a SEM image of hexagonal microcrystals measured in this work with the corresponding TEM spectrum on the right.

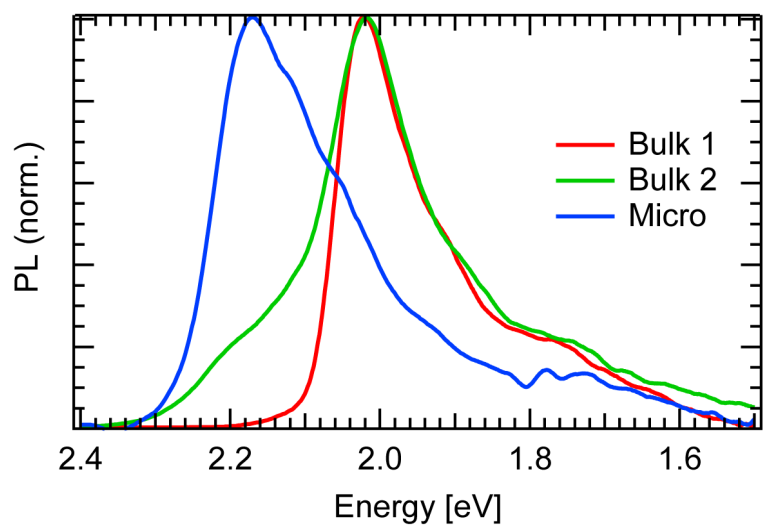

**Figure 6.** Comparison of the integrated PL spectrum of the microcrystals with the corresponding spectra of two bulk crystals with different surface roughness. The effective shift of the spectral weight of the microstructures is due to an increased relative contribution of the high-energy PL peak, rather than to a blueshift.

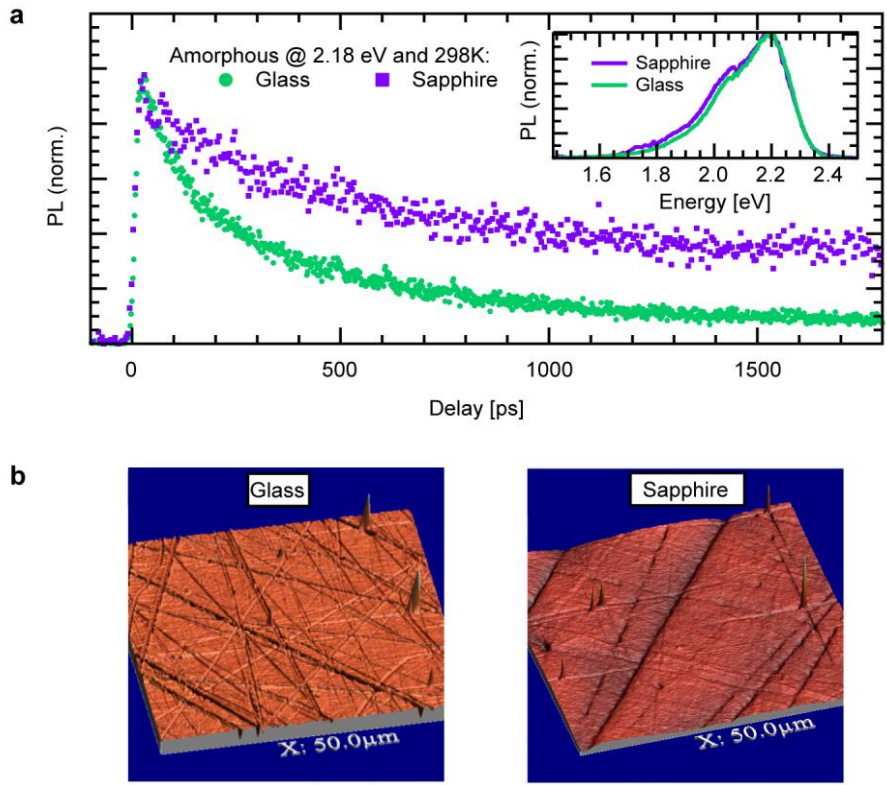

**Figure 7. a** PL-transients of an amorphous film of rubrene on glass and sapphire taken at 2.18 eV at room temperature and the corresponding integrated PL spectra (inset). While the spectra are almost identical, the decay dynamics are considerably different. **b** AFM surface scans of the two amorphous films indicating topographical differences between the two samples. While the film on sapphire exhibits only smooth plateaus, the surface of the film grown on glass is rugged by grain boundaries and dislocations.